\documentclass[
 reprint,
superscriptaddress,
 amsmath,amssymb,
 aps,
pra,
aip,
rsi,]{revtex4-1}
\usepackage{color}
\usepackage{graphicx}% Include figure files
\usepackage{caption}
\usepackage{subcaption}
\usepackage{dcolumn}% Align table columns on decimal point
\usepackage{bm}% bold math
\usepackage{braket}% braket quantum symbols
\usepackage{float}% fix figure position
\usepackage{multirow}
\usepackage{hhline}
\usepackage{mathtools}
\usepackage{qcircuit}
\usepackage{lipsum}

\begin{document}

%\preprint{APS/123-QED}

\title{Extended flag gadgets for low-overhead circuit verification}
\author{Dripto M. Debroy}
\email{dripto@phy.duke.edu}
\affiliation{Department of Physics, Duke University, Durham, NC 27708, USA}

\author{Kenneth R. Brown}
\email{ken.brown@duke.edu}
\affiliation{Department of Physics, Duke University, Durham, NC 27708, USA}
\affiliation{Department of Electrical and Computer Engineering, Duke University, Durham, NC 27708, USA}
\affiliation{Department of Chemistry, Duke University, Durham, NC 27708, USA}
\begin{abstract}
 Flag verification techniques are useful in quantum error correction for detecting critical faults. Here we present an application of flag verification techniques to improving post-selected performance of near-term algorithms. We extend the definition of what constitutes a flag by creating error-detection gadgets based on known transformations of unitary operators. In the case of Clifford or near-Clifford circuits, these unitary operators can be chosen to be controlled Pauli gates, leading to gadgets which require only a small number of additional Clifford gates. We show that such flags can improve circuit fidelities by up to a factor of $2$ after post selection, and demonstrate their effectiveness over error models featuring single-qubit depolarizing noise, crosstalk, and two-qubit coherent overrotation.

\end{abstract}

\pacs{Valid PACS appear here}% PACS, the Physics and Astronomy
                             % Classification Scheme.
%\keywords{Suggested keywords}%Use showkeys class option if keyword
                              %display desired
\maketitle

\section{Introduction}\label{sec:intro} 
In current quantum computers, qubit and gate counts are both at a premium due to system size and gate error limitations. Algorithms which will be productive in this regime cannot require fault-tolerant quantum computation, as even the smallest quantum error correcting proposals require seven to fifteen physical qubits per logical qubit, once fault-tolerant syndrome extraction is taken into account~\cite{steane1996multiple, Laflamme5qubit1996, chao2018quantum, reichardt2020fault, debroy2020logical}. An obvious question is whether there are lower overhead solutions which might help for these algorithms, such as the sampling problems which many expect to be the first promising quantum applications~\cite{peruzzo2014variational, farhi2014quantum, arute2019quantum, AaronsonBosonSample2011}. Some low overhead error mitigation techniques have been considered, like error extrapolation~\cite{Temme_2017, Kandala_2019}, random compiling~\cite{Wallman_2016, Temme_2017, cai2019constructing}, and coherent cancellation~\cite{viola1999dynamical, debroy2018slicing, cai2020mitigating}.

We propose an additional technique, which is an extension of the flag qubit framework~\cite{Yoder2017surfacecodetwist, chao2018quantum, reichardt2020fault, chao2018fault, chamberland2018flag, chao2019flag}. Flag qubits have been employed in the field of quantum error correction to catch critical faults which damage the distance of the code during stabilizer measurement~\cite{li2017fault, chamberland2020topological, chamberland2020triangular}, and for low-overhead error detection during magic state preparation~\cite{chamberland2019fault}. In these cases the flag consists of a small entangle-evolve-disentangle-measure gadget in which the gates that make up the flag commute with the circuit being flagged. In our construction, the flag does not necessarily commute with the circuit but instead has a known propagation through the circuit. We build our flags around verifying that this evolution has occurred correctly. If such a check fails, we know that the circuit application has had an error and it should be discarded. In this way we are able to postselect on these flag outcomes and improve our confidence in the samples we accept. Through this we may be able to reduce the number of experimental runs needed to yield high accuracy results. Our methods supplement existing work on circuits that detect mid-circuit failures through quantum run-time assertions~\cite{liu2020quantum}.

This paper will be organized as follows: In Section~\ref{sec:intro} we introduced the problem, in Section~\ref{sec:NISQflags} we explain the general scheme to extend flag gadgets to larger classes of circuits, in Section~\ref{sec:results} we discuss the performance of our flags in simulation against a spectrum of error models, and in Section~\ref{sec:conclusion} we conclude.
\section{Extending flags to verification of circuits}\label{sec:NISQflags}
\begin{figure}
    \centering
    % \mbox{
    %     \Qcircuit @C=1.3em @R=.7em {
    %          & \qw & \qw & \qw & \qw & \qw & \qw & \targ & \qw & \qw \\
    %          & \qw & \qw & \qw & \qw & \targ & \qw & \qw & \qw & \qw \\
    %          & \qw & \qw & \qw & \targ & \qw & \qw & \qw & \qw & \qw \\
    %          & \qw & \targ & \qw & \qw & \qw & \qw & \qw & \qw & \qw \\ \\
    %         \lstick{|0\rangle} & \gate{H} & \ctrl{-2} & \ctrl{1} & \ctrl{-3} & \ctrl{-4} & \ctrl{1} & \ctrl{-5}  & \gate{H} & \meter\\ 
    %         & & \lstick{|0\rangle} & \targ & \qw & \qw & \targ & \meter & & \\
    %     }
    % }
    \includegraphics[width = 0.95\linewidth]{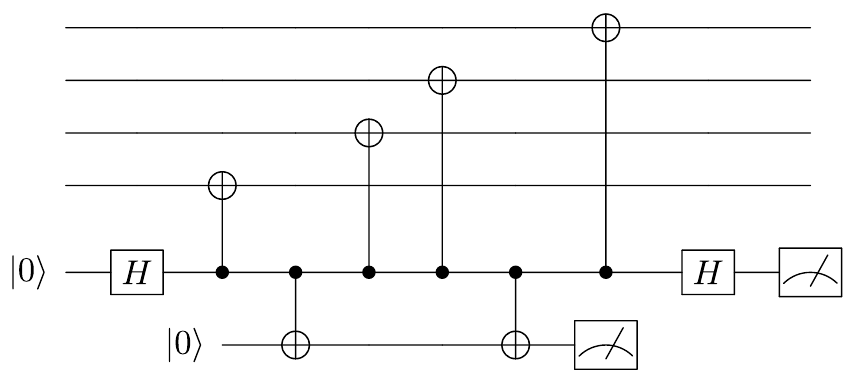}
    \caption{A flag gadget found in \cite{chao2018quantum} which catches $X$-type errors on the ancilla qubit during a syndrome extraction circuit. This prevents faults on the ancilla line from propagating dangerously to the data qubits.}
    \label{fig:basic flag}
    
\end{figure}
\begin{figure*}[ht!]
    \centering
    \includegraphics[width = 0.95\linewidth]{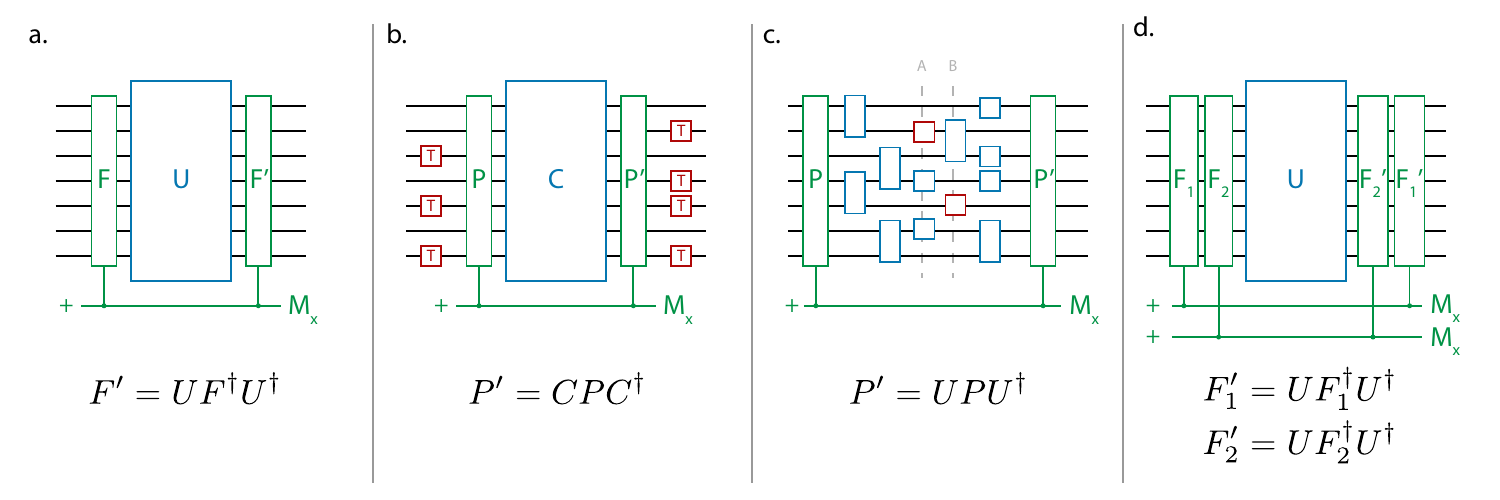}
    \caption{Circuits demonstrating flags for multiple cases. In (a.) we show a general flag gadget for a general $n$-qubit unitary $U$. In (b.) we show that for a Clifford subcircuit (here shown surrounded by $T$-gates) this can be simplified to Pauli flags. In (c.) we show how Pauli flags can be used for a circuit with non-Clifford elements. Lastly in (d.) we show how multiple flags can be used to verify a single circuit.}
    \label{fig:flags}
\end{figure*}
In syndrome extraction circuits, there are often specific fault locations which can propagate dangerously to the data. Flag gadgets verify that these specific faults have not occurred. In the case the fault does occur, the flag gives the decoder enough information to correctly identify and correct the error. A standard flag gadget is shown in Fig.~\ref{fig:basic flag}. In this case, the critical error being caught is an error after the second CNOT gate in the stabilizer measurement. If an $X$-type error on the ancilla occurs after this gate, it would propagate to two errors on the data. Depending on the code being implemented, this might lead to a distance drop. When the flag is implemented, it can detect that this error may have occurred. The stabilizers can then be remeasured and, using this additional piece of information, the code can catch these possible correlated errors. In the standard setup, the entangling and disentangling operators are identical and commute with the circuit.

Our main contribution is extending this framework to the case where the entangling operation does not commute with the circuit (Fig.~\ref{fig:flags}a). Given an n-qubit operator $F$ and an n-qubit unitary $U$, we construct flag gadgets using a $|+\rangle$ state, a controlled-$F$ operator, and a controlled-$F'$ where
\begin{equation}
    F' = U F^{\dagger} U^\dagger.
\end{equation}
If the circuit execution is free of error, $F$ will propagate to $UFU^\dagger$ after the circuit $U$. By defining $F'$ as the inverse of this conjugated operator, the entangling and disentangling flags will cancel out and the original circuits logic is unchanged. An error in $U$ that results in $F$ not propagating to $F'^\dagger$ will prevent perfect disentanglement and become a detectable error on the flag qubit when measured in the $X$ basis. 

Whether postselection on this flag actually improves performance is determined by the number of errors detected by the flag relative to the number of errors added by the flag. Here we construct low-overhead flag gadgets that can verify large circuit sections. As a result, postselection can lead to significant reduction in error rates, increasing confidence in circuit executions which pass all tests.\vspace{-3.5ex}
\subsection{Pauli flags for Clifford circuits}\label{sec:cliffordFlags}
One way to limit the overhead cost of the flag gadget is to limit our entangling operator to a controlled Pauli operator, and the circuit to a Clifford circuit. Since Clifford operators map Pauli operators to other Pauli operators, the disentangling operator is guaranteed to also be a controlled Pauli operator. As a result we can be assured that our flag gadget is composed of at most $2n$ two-qubit gates, where n is the number of qubits in the unflagged circuit (Fig.~\ref{fig:flags}b).

\begin{figure}[h!]
    \centering
    % \mbox{
    %     \Qcircuit @C=1.2em @R=.7em {
    %         & \qw & \gate{T} & \gate{Z} & \gate{Z} & \qw & \ctrl{4} & \gate{Z} & \gate{H} & \meter\\
    %         & \qw & \gate{T} & \qw & \gate{Z} \qwx[-1] & \ctrl{3} & \qw & \qw & \gate{H} & \meter\\
    %         & \qw & \gate{T} & \gate{Z} \qwx[-2] & \ctrl{-1} \qwx[2] & \qw & \qw & \qw & \gate{H} & \meter\\
    %         & \qw & \gate{T} & \ctrl{-1} \qwx[1] & \qw & \qw & \qw & \gate{Z} & \gate{H} & \meter\\
    %         & \qw & \gate{T} & \gate{XZ} & \gate{X} & \gate{X} & \gate{XZ} & \gate{Z} & \gate{H} & \gate{Y} & \qw\\
    %     }
    % }
    \includegraphics[width = \linewidth]{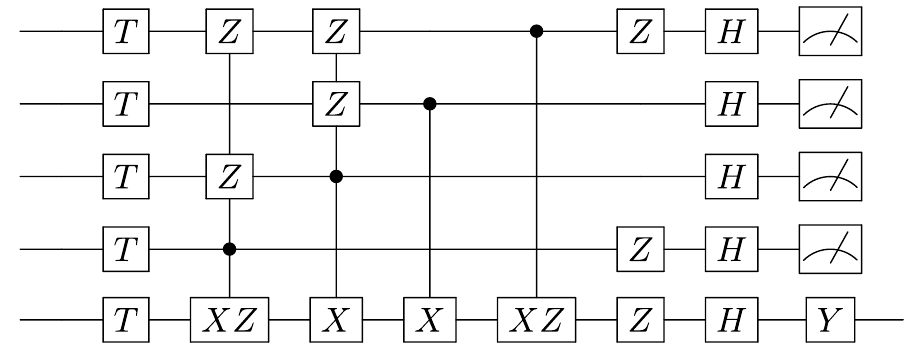}
    \caption{Circuit from \cite{jochym2012robustness} which is used to improve the fidelity of magic states. The Clifford block after the T-gate layer can be flagged by a Pauli flag following Fig.~\ref{fig:flags}b. We do not consider the measurements at the end of the circuit in order to make our results more general. Flags accounting for these measurements could focus on catching only errors which flip the outcome, further optimizing performance.}
    \label{fig:magic circuit}
\end{figure}

While a quantum algorithm requires non-Clifford elements in order to be useful, one could imagine flagging sections of a larger circuit in order to catch errors in each section. In Section~\ref{sec: magic results} we present the performance of Clifford flags when applied to the circuit seen in Fig.~\ref{fig:magic circuit}.
\subsection{Pauli flags for circuits with low non-Clifford gate density}
A circuit which has few non-Clifford elements can also be flagged with Pauli operators in some cases, however more care must be put into finding the appropriate Pauli operators. Specifically, in order to avoid having to consider Pauli propagation through non-Clifford gates, it would be ideal for the flags to never interact with these gates. 

\begin{figure}[t!]
    \centering
    % \mbox{
    %     \Qcircuit @C=1.2em @R=.7em {
    %         & \qw & \ctrl{1} & \qw & \qw & \qw & \qw & \qw & \qw & \qw & \ctrl{1} & \qw\\
    %         & \qw & \targ & \ctrl{1} & \qw & \qw & \qw & \qw & \qw & \ctrl{1} & \targ & \qw\\
    %         & \qw & \qw & \targ & \ctrl{1} & \qw & \qw & \qw & \ctrl{1} & \targ & \qw & \qw\\
    %         & \qw & \qw & \qw & \targ & \ctrl{1} & \qw & \ctrl{1} & \targ & \qw & \qw & \qw\\
    %         & \qw & \qw & \qw & \qw & \targ & \gate{Z(\theta)} & \targ & \qw & \qw & \qw & \qw
    %     }
    % }
    \includegraphics[width = 0.95\linewidth]{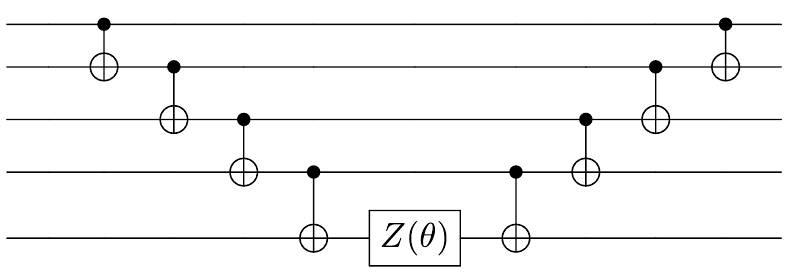}
    \caption{A circuit for $ZZZZZ(\theta)$ that has a single non-Clifford element. This circuit can still be flagged with Pauli operators by designing the flags to avoid or commute with the non-Clifford gate, as shown in Fig.~\ref{fig:flags}c.}
    \label{fig:ZZZZZ}
\end{figure}

The goal is to find a Pauli operator $P$ such that, as it is propagated layer-by-layer through the circuit, it acts as an identity operator on all qubit lines that have a non-Clifford gate, as shown in Fig.~\ref{fig:flags}c. In the figure, this class of entangling flags $P$ is constrained by the fact that as you propagate $P$ through to layer A, its action must commute with the non-Clifford gate on qubit 2, and on layer B, the propagated Pauli operator must commute with the non-Clifford gate on qubit 5. This could either mean that the Pauli operator on those layers acts as identity on that qubit, or if the red gates are arbitrary angle Pauli rotations, the propagated Pauli operator must be along the same axis. Finding such flags will get more difficult as the density of non-Clifford gates increases, but for a circuit like the non-Clifford Pauli rotation circuit in Fig.~\ref{fig:ZZZZZ}, such a flag can be found easily by starting with a Pauli operator in the middle of the circuit that avoids the non-Clifford gate and propagating outwards in both directions. We show the value of flags of this type in Section~\ref{sec: ZZZZZ}.
\subsection{Implementing multiple flags on one circuit}
Since the flag leaves the circuit unchanged, multiple flags can be checked on the same circuit execution. The flags must be nested such that the order of entangling flags is reversed when checking the disentangling flags, as shown in Fig.~\ref{fig:flags}d. As long as this condition is respected and each flag is compatible with the circuit, the logic will be unchanged. What we see in our simulations is that in the cases where the flags are doing well on their own, combining them seems to be productive. The data shown in the following section (Sec.~\ref{sec:results}) is representative of this behavior. In error models where the flags are not performing as well on their own, additional flagging does not generally seem to flip that behavior. When designing these flag pairs, it is beneficial to have the flags disagree on most qubits, so that they can catch a wider range of errors. Additionally, the penalty in terms of extra fault locations is somewhat alleviated for the inner flag since some of its faults will be caught by the outer flag.
\subsection{Finding good flags}
\begin{figure}
    \centering
    \includegraphics[trim = 10 0 0 0, width = 0.9\linewidth]{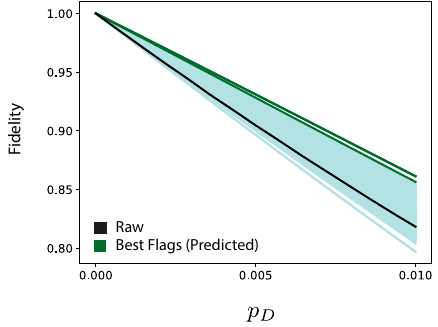}
    \caption{(Color Online) Data verifying that the metric in Eq.~\ref{eq:qual} effectively captures the performance of flags within the magic state distillation circuit in Fig.~\ref{fig:magic circuit}. The fidelities of the flagged circuits are shown in green (gray in print), with the three flags with the highest values of $q(P, P')$ shown in dark green (darker gray). The raw circuit is shown in black. The data shows that the three flags with the highest values of $q(P, P')$ were also the top three performers in terms of fidelity. This metric might not perform as well at higher error rates where the higher order terms in the polynomial are more important.}
    \label{fig:goodflags}
\end{figure}

\begin{figure*}[t!]
    \centering
    \includegraphics[width = 0.95\linewidth]{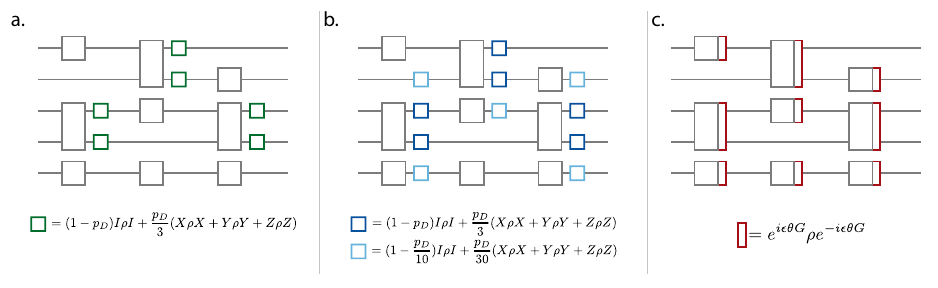}
    \caption{(Color Online) The (a.) single-qubit depolarizing, (b.) single-qubit depolarizing with crosstalk, and (c.) overrotation error models we consider applied to a random circuit. Gray rectangles are the desired gates in the circuit, while colored (smaller) rectangles are errors. In the first two models the errors are depolarizing errors, where in the crosstalk model nearest neighbors experience a $\frac{1}{10}\times$ the gate failure rate. For the overrotation model, every gate is followed by a slight overrotation along the same axis by an amount $\epsilon$.}
    \label{fig:errors}
\end{figure*}
If these flags are to be useful for larger circuits and more general  applications, it is imperative that we are able to predict which flags will be successful without full simulation. To that end, there are a few guidelines which lead to an effective flag:
\begin{itemize}
    \item The circuit being flagged should have a high gate density, meaning many gates on relatively few qubits. Such circuits will have a large number of error locations which can be effectively flagged without adding significant overhead.
    \item If the algorithm ends with measurements, the final flag should match the basis of the measurement. For example if an algorithm ends with measurements on all qubits in the $Z$-basis, then having the final flag be a $Z$-type operator will detect the widest set of errors which flip the results of the measurement. This will also minimize the impact of gate error if the error model is one in which gate error is along the axis of the gate.
    \item If multiple flags are being used, they should be designed so each catches different errors on each qubit, to increase sensitivity to a variety of errors. This also allows for the gate errors from the inner flag to be caught by the outer flag in error models like those discussed in the previous point.
\end{itemize}
On top of this, it is useful to note that since these flags do not lead to schemes which possess a distance, the dominant term in the error polynomial will be the linear term. For a given error model, it is easy to find the effective error model at the output layer of the circuit by propagating single errors through the circuit from each fault location. By checking how many of these errors each flag detects, balanced against the number of extra error locations introduced by the flag, one can efficiently guess which flags in a set will be the highest performing. An example of this method is shown in Fig.~\ref{fig:goodflags} for the single-qubit depolarizing error model. We create a set of all possible Pauli errors on the output of the circuit from a single fault, and define a quality metric on a flag $\{P, P'\}$:
\begin{equation}\label{eq:qual}
    q(P, P') = N_{detected}  - 6w(P) - 6w(P'),
\end{equation}
where $N_{detected}$ is the number of faults in the output error set that anticommute with $P'$, and $w(P)$ is the weight of a Pauli operator $P$. The weights are multiplied by $6$ since each two-qubit gate adds $6$ possible faults to the system, $3$ on the control and $3$ on the target, under the error model described in Fig.~\ref{fig:errors}a. We calculated this metric and found that the flags which had the three largest values of $q(P, P')$ also performed the best in terms of fidelity. This supports the effectiveness of this quality metric in Eq.~\ref{eq:qual}, and calculating this metric is polynomial in the number of gates in the circuit. We note that this scheme will work for any error model, although the coefficients on $w(P)$ and $w(P')$ must be updated to properly represent the number of fault locations introduced by the flag gadget.
\section{Results}\label{sec:results}
\begin{figure*}
    \centering
    \includegraphics[width = 1\linewidth]{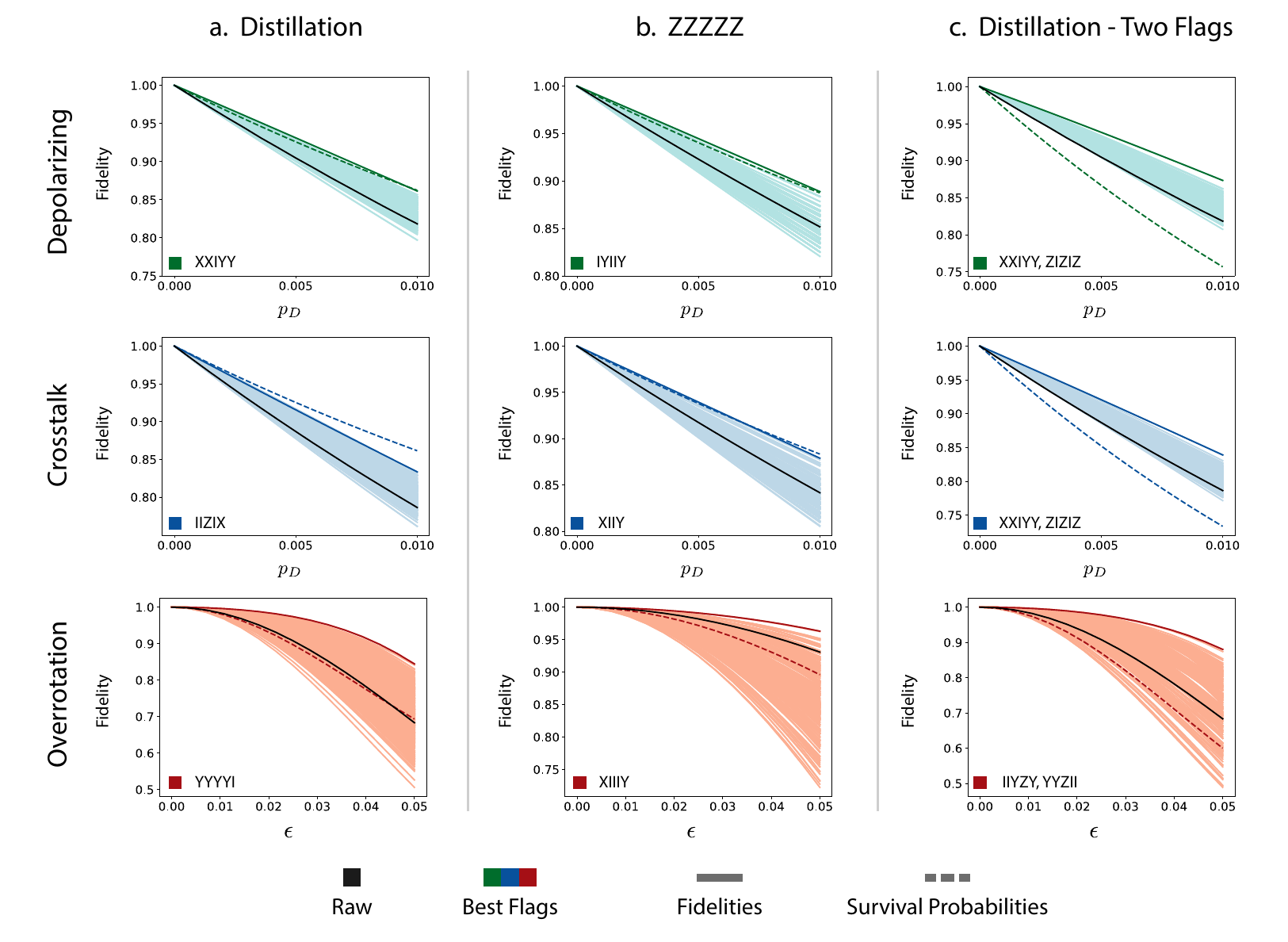}
    \caption{(Color Online) Data for flagging the (a.) magic state distillation circuit in Fig.~\ref{fig:magic circuit}, the (b.) $ZZZZZ(\theta)$ circuit in Fig.~\ref{fig:ZZZZZ}, and the (c.) magic state distillation circuit with two flag gadgets. The data for each error model is colored to match Fig.~\ref{fig:errors}, where the best performing flag has both its fidelity (solid line) and survival probability (dashed line) shown darker, and is labeled within the figure by its disentangling operator. Raw circuit fidelities are shown in black. Due to being higher in gate density, the magic state distillation circuit is more appropriate for flagging. Similarly, the multiple flag solutions work better in the single-qubit depolarizing and crosstalk error models, which lend themselves better to flagging as well.}
    \label{fig:data}
\end{figure*}

All results in this section are acquired by density matrix simulations written using Cirq~\cite{cirq}. Data is shown for 500 Pauli flags that were randomly generated for each circuit. It should be noted that this is a computationally efficient procedure~\cite{aaronson2004improved}. The data for multiple flag systems are created using 100 pairs of randomly generated flags. Fidelities are taken before any measurements in the initial circuit, and defined by 
\begin{equation*}
    F(\rho_{noisy}, |\psi_{clean}\rangle) = \langle \psi_{clean}|\rho_{noisy}|\psi_{clean}\rangle.
\end{equation*}
Survival probabilities are presented for the best performing flag, and are defined as the likelihood of the flag qubit measuring $+1$. As a result, flags which detect the most errors will have the worst survival probabilities. This tradeoff is worthwhile in algorithms for which an error leads to an outcome which is no longer valuable.
\vspace{-1.5ex}
\subsection{Error Models}\label{sec:error models}
In this work we consider a set of three error models, shown in Fig.~\ref{fig:errors}, which are pulled from the spectrum of errors seen in physical systems today. 

Firstly, we consider a single-qubit depolarizing model where single-qubit depolarizing errors follow all two-qubit gates. This model has been previously looked at by the Google team in \cite{huggins2019efficient}. Secondly, we consider a simplified crosstalk model which takes this single-qubit depolarizing model and adds weak depolarizing noise to the qubits adjacent to any two-qubit gate. Lastly, we consider a coherent overrotation error model based on ion trap quantum computers. These errors have also been studied in previous works~\cite{debroy2018slicing, debroy2020logical}, and the data presented for them are for circuits compiled down to ion trap gates~\cite{MaslovCircuitCompIT2017}. Through this set of models we show that our techniques succeed against a wide variety of errors seen in physical hardware.
\subsection{Clifford Example: Magic State Distillation}\label{sec: magic results}
In Fig.~\ref{fig:data}a we show the performance of 500 randomly generated flags on the circuit in Fig.~\ref{fig:magic circuit}. As this circuit is densely populated with gates, a majority of the flags succeed in improving the performance of the circuit. The depolarizing and crosstalk models are much easier to flag, as they give off discrete errors which can be caught by the flags. The coherent gate error model is more damaging in general, and also can easily create error patterns where multiple errors anticommute with the flag at the same time. Since the flag will only detect an error when an odd number of errors anticommute with it, there will still be unitary errors on the system after measurement. Additionally any coherent error model can have overrotations which combine coherently, so the gate error may constructively interfere with our flags themselves.

The wider spread on the coherent case is because that error model creates the most structured error model on the output. The depolarizing and crosstalk channels will have some asymmetry on the output errors, due to the fact that errors can only propagate in certain ways. However in the coherent overrotation case, the asymmetry on the gate errors themselves also carries over directly to the output. As a result, flags have a wider spread under this error model, in both positive and negative contributions. Due to the extremely structured error model, systems that exhibit overrotation errors would be able to strongly benefit from flags that are specifically designed to take advantage of this error asymmetry.
\subsection{Non-Clifford Example: Arbitrary Angle Pauli Rotations}\label{sec: ZZZZZ}
In Fig.~\ref{fig:data}b we present data from 500 flags on the $ZZZZZ(\theta)$ circuit in Fig.~\ref{fig:ZZZZZ}. This circuit is significantly less dense than the magic state distillation circuit in Fig.~\ref{fig:magic circuit}, and as a result the flags do not perform as well on average. This is especially true in the case of the coherent overrotation model, as this circuit's structure propagates errors in damaging ways. If a circuit would require consecutive multi-qubit Pauli rotations, the product of all of them could be flagged instead. This would lead to a better tradeoff between error locations added and subtracted from the system, at the expense of missing some error patterns which would be caught by a scheme which flagged each rotation individually.
\subsection{Multiple Flag Results}\label{sec: 2flag}
In Fig.~\ref{fig:data}c we show how pairs of Pauli flags can perform on the magic state distillation circuit in Fig.~\ref{fig:magic circuit}. The pairs of flags behave as one would expect, in cases where the initial flags were performing well, they improve things further. Specifically, in the single-flag simulations for the depolarizing and crosstalk models, there were a good number of flags which performed worse than the unflagged case. In the two-flag case, these two error models almost always saw improvements over the raw circuit. In the overrotation error model the main impact was an increased spread in the outcomes for different flags. This is due to the fact that a coherent error model can see significantly worsened performance when the gate errors from the flag coherently add with the gate errors from the circuit, as opposed to a depolarizing case where a poorly chosen flag adds the same amount of error as a well chosen one, but simply detects fewer errors.
\section{Conclusion}\label{sec:conclusion}
In this work we have presented a method for applying flag verification techniques to error detection in Clifford and near-Clifford circuits. These flags can be efficiently computed and ranked, meaning that even circuits that are outside of the range of full simulation can be protected. The flagged circuits show improvements in performance when postselected for flag success, even in cases where there are relatively few gates. Additionally, the survival probabilities for these techniques are high enough relative to the performance boosts they provide such that postselecting on them should improve performance in sampling problems. There is also a potential value in investing resources into finding optimal flags for common Clifford subroutines which occur often within algorithms. 

Recent work on understanding the Clifford hierarchy may offer interesting non-Clifford operators that possess guarantees on the existence of commuting Pauli operators~\cite{pllaha2020unweyling}. In situations where there exist known propagations for certain operators through a relevant non-Clifford block, flags of the type shown in this paper should provide similar improvements. Through these techniques, this idea could be extended beyond Clifford and near-Clifford circuits. These techniques may be invaluable in near-term experiments where the overhead costs of more effective error correction techniques are prohibitive.

\section{Acknowledgements}
The authors thank Jarrod McClean and Cody Jones for helpful conversations. This work was supported by the National Science Foundation (NSF) STAQ project (1818914) and EPiQC --- an NSF Expedition in Computing (1730104). D.M.D. is also funded in part by an NSF QISE-NET fellowship (1747426).
 
% \bibliographystyle{apsrev}
% \bibliography{References}

\begin{thebibliography}{31}
\expandafter\ifx\csname natexlab\endcsname\relax\def\natexlab#1{#1}\fi
\expandafter\ifx\csname bibnamefont\endcsname\relax
  \def\bibnamefont#1{#1}\fi
\expandafter\ifx\csname bibfnamefont\endcsname\relax
  \def\bibfnamefont#1{#1}\fi
\expandafter\ifx\csname citenamefont\endcsname\relax
  \def\citenamefont#1{#1}\fi
\expandafter\ifx\csname url\endcsname\relax
  \def\url#1{\texttt{#1}}\fi
\expandafter\ifx\csname urlprefix\endcsname\relax\def\urlprefix{URL }\fi
\providecommand{\bibinfo}[2]{#2}
\providecommand{\eprint}[2][]{\url{#2}}

\bibitem[{\citenamefont{Steane}(1996)}]{steane1996multiple}
\bibinfo{author}{\bibfnamefont{A.}~\bibnamefont{Steane}},
  \bibinfo{journal}{Proc. of the Royal Soc. of London. Series A: Math., Phys.
  and Eng. Sciences} \textbf{\bibinfo{volume}{452}}, \bibinfo{pages}{2551}
  (\bibinfo{year}{1996}).

\bibitem[{\citenamefont{Laflamme et~al.}(1996)\citenamefont{Laflamme, Miquel,
  Paz, and Zurek}}]{Laflamme5qubit1996}
\bibinfo{author}{\bibfnamefont{R.}~\bibnamefont{Laflamme}},
  \bibinfo{author}{\bibfnamefont{C.}~\bibnamefont{Miquel}},
  \bibinfo{author}{\bibfnamefont{J.~P.} \bibnamefont{Paz}}, \bibnamefont{and}
  \bibinfo{author}{\bibfnamefont{W.~H.} \bibnamefont{Zurek}},
  \bibinfo{journal}{Phys. Rev. Lett.} \textbf{\bibinfo{volume}{77}},
  \bibinfo{pages}{198} (\bibinfo{year}{1996}).

\bibitem[{\citenamefont{Chao and
  Reichardt}(2018{\natexlab{a}})}]{chao2018quantum}
\bibinfo{author}{\bibfnamefont{R.}~\bibnamefont{Chao}} \bibnamefont{and}
  \bibinfo{author}{\bibfnamefont{B.~W.} \bibnamefont{Reichardt}},
  \bibinfo{journal}{Phys. Rev. Lett.} \textbf{\bibinfo{volume}{121}},
  \bibinfo{pages}{050502} (\bibinfo{year}{2018}{\natexlab{a}}).
  
  \bibitem[{\citenamefont{Reichardt}(2020)}]{reichardt2020fault}
  \bibinfo{author}{\bibfnamefont{B.~W.} \bibnamefont{Reichardt}},
  \bibinfo{journal}{Quantum Sci. and Tech.} \textbf{\bibinfo{volume}{6}},
  \bibinfo{pages}{015007} (\bibinfo{year}{2020}).

  
\bibitem[{\citenamefont{Debroy et~al.}(2020)\citenamefont{Debroy, Li, Huang,
  and Brown}}]{debroy2020logical}
\bibinfo{author}{\bibfnamefont{D.~M.} \bibnamefont{Debroy}},
  \bibinfo{author}{\bibfnamefont{M.}~\bibnamefont{Li}},
  \bibinfo{author}{\bibfnamefont{S.}~\bibnamefont{Huang}}, \bibnamefont{and}
  \bibinfo{author}{\bibfnamefont{K.~R.} \bibnamefont{Brown}},
  \bibinfo{journal}{Quantum Science and Technology}
  \textbf{\bibinfo{volume}{5}}, \bibinfo{pages}{034002} (\bibinfo{year}{2020}).

\bibitem[{\citenamefont{Peruzzo et~al.}(2014)\citenamefont{Peruzzo, McClean,
  Shadbolt, Yung, Zhou, Love, Aspuru-Guzik, and
  O’brien}}]{peruzzo2014variational}
\bibinfo{author}{\bibfnamefont{A.}~\bibnamefont{Peruzzo}},
  \bibinfo{author}{\bibfnamefont{J.}~\bibnamefont{McClean}},
  \bibinfo{author}{\bibfnamefont{P.}~\bibnamefont{Shadbolt}},
  \bibinfo{author}{\bibfnamefont{M.-H.} \bibnamefont{Yung}},
  \bibinfo{author}{\bibfnamefont{X.-Q.} \bibnamefont{Zhou}},
  \bibinfo{author}{\bibfnamefont{P.~J.} \bibnamefont{Love}},
  \bibinfo{author}{\bibfnamefont{A.}~\bibnamefont{Aspuru-Guzik}},
  \bibnamefont{and} \bibinfo{author}{\bibfnamefont{J.~L.}
  \bibnamefont{O’brien}}, \bibinfo{journal}{Nature Comm.}
  \textbf{\bibinfo{volume}{5}}, \bibinfo{pages}{4213} (\bibinfo{year}{2014}).

\bibitem[{\citenamefont{Farhi et~al.}(2014)\citenamefont{Farhi, Goldstone, and
  Gutmann}}]{farhi2014quantum}
\bibinfo{author}{\bibfnamefont{E.}~\bibnamefont{Farhi}},
  \bibinfo{author}{\bibfnamefont{J.}~\bibnamefont{Goldstone}},
  \bibnamefont{and} \bibinfo{author}{\bibfnamefont{S.}~\bibnamefont{Gutmann}},
  \bibinfo{journal}{arXiv preprint arXiv:1411.4028}  (\bibinfo{year}{2014}).

\bibitem[{\citenamefont{Arute et~al.}(2019)\citenamefont{Arute, Arya, Babbush,
  Bacon, Bardin, Barends, Biswas, Boixo, Brandao, Buell
  et~al.}}]{arute2019quantum}
\bibinfo{author}{\bibfnamefont{F.}~\bibnamefont{Arute}},
  \bibinfo{author}{\bibfnamefont{K.}~\bibnamefont{Arya}},
  \bibinfo{author}{\bibfnamefont{R.}~\bibnamefont{Babbush}},
  \bibinfo{author}{\bibfnamefont{D.}~\bibnamefont{Bacon}},
  \bibinfo{author}{\bibfnamefont{J.~C.} \bibnamefont{Bardin}},
  \bibinfo{author}{\bibfnamefont{R.}~\bibnamefont{Barends}},
  \bibinfo{author}{\bibfnamefont{R.}~\bibnamefont{Biswas}},
  \bibinfo{author}{\bibfnamefont{S.}~\bibnamefont{Boixo}},
  \bibinfo{author}{\bibfnamefont{F.~G.} \bibnamefont{Brandao}},
  \bibinfo{author}{\bibfnamefont{D.~A.} \bibnamefont{Buell}},
  \bibnamefont{et~al.}, \bibinfo{journal}{Nature}
  \textbf{\bibinfo{volume}{574}}, \bibinfo{pages}{505} (\bibinfo{year}{2019}).

\bibitem[{\citenamefont{Aaronson and Arkhipov}(2011)}]{AaronsonBosonSample2011}
\bibinfo{author}{\bibfnamefont{S.}~\bibnamefont{Aaronson}} \bibnamefont{and}
  \bibinfo{author}{\bibfnamefont{A.}~\bibnamefont{Arkhipov}}
  (\bibinfo{year}{2011}), STOC, pp. \bibinfo{pages}{333--342}.

\bibitem[{\citenamefont{Temme et~al.}(2017)\citenamefont{Temme, Bravyi, and
  Gambetta}}]{Temme_2017}
\bibinfo{author}{\bibfnamefont{K.}~\bibnamefont{Temme}},
  \bibinfo{author}{\bibfnamefont{S.}~\bibnamefont{Bravyi}}, \bibnamefont{and}
  \bibinfo{author}{\bibfnamefont{J.~M.} \bibnamefont{Gambetta}},
  \bibinfo{journal}{Phys. Rev. Lett.} \textbf{\bibinfo{volume}{119}}
  (\bibinfo{year}{2017}).

\bibitem[{\citenamefont{Kandala et~al.}(2019)\citenamefont{Kandala, Temme,
  Córcoles, Mezzacapo, Chow, and Gambetta}}]{Kandala_2019}
\bibinfo{author}{\bibfnamefont{A.}~\bibnamefont{Kandala}},
  \bibinfo{author}{\bibfnamefont{K.}~\bibnamefont{Temme}},
  \bibinfo{author}{\bibfnamefont{A.~D.} \bibnamefont{Córcoles}},
  \bibinfo{author}{\bibfnamefont{A.}~\bibnamefont{Mezzacapo}},
  \bibinfo{author}{\bibfnamefont{J.~M.} \bibnamefont{Chow}}, \bibnamefont{and}
  \bibinfo{author}{\bibfnamefont{J.~M.} \bibnamefont{Gambetta}},
  \bibinfo{journal}{Nature} \textbf{\bibinfo{volume}{567}},
  \bibinfo{pages}{491–495} (\bibinfo{year}{2019}).

\bibitem[{\citenamefont{Wallman and Emerson}(2016)}]{Wallman_2016}
\bibinfo{author}{\bibfnamefont{J.~J.} \bibnamefont{Wallman}} \bibnamefont{and}
  \bibinfo{author}{\bibfnamefont{J.}~\bibnamefont{Emerson}},
  \bibinfo{journal}{Phys. Rev. A} \textbf{\bibinfo{volume}{94}}
  (\bibinfo{year}{2016}).

\bibitem[{\citenamefont{Cai and Benjamin}(2019)}]{cai2019constructing}
\bibinfo{author}{\bibfnamefont{Z.}~\bibnamefont{Cai}} \bibnamefont{and}
  \bibinfo{author}{\bibfnamefont{S.~C.} \bibnamefont{Benjamin}},
  \bibinfo{journal}{Scientific Reports} \textbf{\bibinfo{volume}{9}},
  \bibinfo{pages}{1} (\bibinfo{year}{2019}).

\bibitem[{\citenamefont{Viola et~al.}(1999)\citenamefont{Viola, Knill, and
  Lloyd}}]{viola1999dynamical}
\bibinfo{author}{\bibfnamefont{L.}~\bibnamefont{Viola}},
  \bibinfo{author}{\bibfnamefont{E.}~\bibnamefont{Knill}}, \bibnamefont{and}
  \bibinfo{author}{\bibfnamefont{S.}~\bibnamefont{Lloyd}},
  \bibinfo{journal}{Phys. Rev. Lett.} \textbf{\bibinfo{volume}{82}},
  \bibinfo{pages}{2417} (\bibinfo{year}{1999}).

\bibitem[{\citenamefont{Debroy et~al.}(2018)\citenamefont{Debroy, Li, Newman,
  and Brown}}]{debroy2018slicing}
\bibinfo{author}{\bibfnamefont{D.~M.} \bibnamefont{Debroy}},
  \bibinfo{author}{\bibfnamefont{M.}~\bibnamefont{Li}},
  \bibinfo{author}{\bibfnamefont{M.}~\bibnamefont{Newman}}, \bibnamefont{and}
  \bibinfo{author}{\bibfnamefont{K.~R.} \bibnamefont{Brown}},
  \bibinfo{journal}{Phys. Rev. Lett.} \textbf{\bibinfo{volume}{121}},
  \bibinfo{pages}{250502} (\bibinfo{year}{2018}).

\bibitem[{\citenamefont{Cai et~al.}(2020)\citenamefont{Cai, Xu, and
  Benjamin}}]{cai2020mitigating}
\bibinfo{author}{\bibfnamefont{Z.}~\bibnamefont{Cai}},
  \bibinfo{author}{\bibfnamefont{X.}~\bibnamefont{Xu}}, \bibnamefont{and}
  \bibinfo{author}{\bibfnamefont{S.~C.} \bibnamefont{Benjamin}},
  \bibinfo{journal}{npj Quantum Information} \textbf{\bibinfo{volume}{6}},
  \bibinfo{pages}{1} (\bibinfo{year}{2020}).

\bibitem[{\citenamefont{Yoder and Kim}(2017)}]{Yoder2017surfacecodetwist}
\bibinfo{author}{\bibfnamefont{T.~J.} \bibnamefont{Yoder}} \bibnamefont{and}
  \bibinfo{author}{\bibfnamefont{I.~H.} \bibnamefont{Kim}},
  \bibinfo{journal}{{Quantum}} \textbf{\bibinfo{volume}{1}}, \bibinfo{pages}{2}
  (\bibinfo{year}{2017}).

\bibitem[{\citenamefont{Chao and
  Reichardt}(2018{\natexlab{b}})}]{chao2018fault}
\bibinfo{author}{\bibfnamefont{R.}~\bibnamefont{Chao}} \bibnamefont{and}
  \bibinfo{author}{\bibfnamefont{B.~W.} \bibnamefont{Reichardt}},
  \bibinfo{journal}{npj Quantum Information} \textbf{\bibinfo{volume}{4}},
  \bibinfo{number}{42} (\bibinfo{year}{2018}{\natexlab{b}}).

\bibitem[{\citenamefont{Chamberland and Beverland}(2018)}]{chamberland2018flag}
\bibinfo{author}{\bibfnamefont{C.}~\bibnamefont{Chamberland}} \bibnamefont{and}
  \bibinfo{author}{\bibfnamefont{M.~E.} \bibnamefont{Beverland}},
  \bibinfo{journal}{Quantum} \textbf{\bibinfo{volume}{2}}, \bibinfo{pages}{10}
  (\bibinfo{year}{2018}).

\bibitem[{\citenamefont{Chao and Reichardt}(2019)}]{chao2019flag}
\bibinfo{author}{\bibfnamefont{R.}~\bibnamefont{Chao}} \bibnamefont{and}
  \bibinfo{author}{\bibfnamefont{B.~W.} \bibnamefont{Reichardt}},
  \bibinfo{journal}{arXiv preprint arXiv:1912.09549}  (\bibinfo{year}{2019}).

\bibitem[{\citenamefont{Li et~al.}(2017)\citenamefont{Li, Guti{\'e}rrez, David,
  Hernandez, and Brown}}]{li2017fault}
\bibinfo{author}{\bibfnamefont{M.}~\bibnamefont{Li}},
  \bibinfo{author}{\bibfnamefont{M.}~\bibnamefont{Guti{\'e}rrez}},
  \bibinfo{author}{\bibfnamefont{S.~E.} \bibnamefont{David}},
  \bibinfo{author}{\bibfnamefont{A.}~\bibnamefont{Hernandez}},
  \bibnamefont{and} \bibinfo{author}{\bibfnamefont{K.~R.} \bibnamefont{Brown}},
  \bibinfo{journal}{Phys. Rev. A} \textbf{\bibinfo{volume}{96}},
  \bibinfo{pages}{032341} (\bibinfo{year}{2017}).

\bibitem[{\citenamefont{Chamberland
  et~al.}(2020{\natexlab{a}})\citenamefont{Chamberland, Zhu, Yoder, Hertzberg,
  and Cross}}]{chamberland2020topological}
\bibinfo{author}{\bibfnamefont{C.}~\bibnamefont{Chamberland}},
  \bibinfo{author}{\bibfnamefont{G.}~\bibnamefont{Zhu}},
  \bibinfo{author}{\bibfnamefont{T.~J.} \bibnamefont{Yoder}},
  \bibinfo{author}{\bibfnamefont{J.~B.} \bibnamefont{Hertzberg}},
  \bibnamefont{and} \bibinfo{author}{\bibfnamefont{A.~W.} \bibnamefont{Cross}},
  \bibinfo{journal}{Phys. Rev. X} \textbf{\bibinfo{volume}{10}},
  \bibinfo{pages}{011022} (\bibinfo{year}{2020}{\natexlab{a}}).

\bibitem[{\citenamefont{Chamberland
  et~al.}(2020{\natexlab{b}})\citenamefont{Chamberland, Kubica, Yoder, and
  Zhu}}]{chamberland2020triangular}
\bibinfo{author}{\bibfnamefont{C.}~\bibnamefont{Chamberland}},
  \bibinfo{author}{\bibfnamefont{A.}~\bibnamefont{Kubica}},
  \bibinfo{author}{\bibfnamefont{T.~J.} \bibnamefont{Yoder}}, \bibnamefont{and}
  \bibinfo{author}{\bibfnamefont{G.}~\bibnamefont{Zhu}}, \bibinfo{journal}{New
  Journal of Physics} \textbf{\bibinfo{volume}{22}}, \bibinfo{pages}{023019}
  (\bibinfo{year}{2020}{\natexlab{b}}).

\bibitem[{\citenamefont{Chamberland and Cross}(2019)}]{chamberland2019fault}
\bibinfo{author}{\bibfnamefont{C.}~\bibnamefont{Chamberland}} \bibnamefont{and}
  \bibinfo{author}{\bibfnamefont{A.~W.} \bibnamefont{Cross}},
  \bibinfo{journal}{Quantum} \textbf{\bibinfo{volume}{3}}, \bibinfo{pages}{143}
  (\bibinfo{year}{2019}).

\bibitem[{\citenamefont{Liu et~al.}(2020)\citenamefont{Liu, Byrd, and
  Zhou}}]{liu2020quantum}
\bibinfo{author}{\bibfnamefont{J.}~\bibnamefont{Liu}},
  \bibinfo{author}{\bibfnamefont{G.~T.} \bibnamefont{Byrd}}, \bibnamefont{and}
  \bibinfo{author}{\bibfnamefont{H.}~\bibnamefont{Zhou}}, in
  \emph{\bibinfo{booktitle}{ASPLOS}} (\bibinfo{year}{2020}), pp.
  \bibinfo{pages}{1017--1030}.

\bibitem[{\citenamefont{Jochym-O'Connor
  et~al.}(2012)\citenamefont{Jochym-O'Connor, Yu, Helou, and
  Laflamme}}]{jochym2012robustness}
\bibinfo{author}{\bibfnamefont{T.}~\bibnamefont{Jochym-O'Connor}},
  \bibinfo{author}{\bibfnamefont{Y.}~\bibnamefont{Yu}},
  \bibinfo{author}{\bibfnamefont{B.}~\bibnamefont{Helou}}, \bibnamefont{and}
  \bibinfo{author}{\bibfnamefont{R.}~\bibnamefont{Laflamme}},
  \bibinfo{journal}{arXiv preprint arXiv:1205.6715}  (\bibinfo{year}{2012}).

\bibitem[{cir()}]{cirq}
\bibinfo{howpublished}{https://github.com/quantumlib/Cirq}.

\bibitem[{\citenamefont{Aaronson and Gottesman}(2004)}]{aaronson2004improved}
\bibinfo{author}{\bibfnamefont{S.}~\bibnamefont{Aaronson}} \bibnamefont{and}
  \bibinfo{author}{\bibfnamefont{D.}~\bibnamefont{Gottesman}},
  \bibinfo{journal}{Phys. Rev. A} \textbf{\bibinfo{volume}{70}},
  \bibinfo{pages}{052328} (\bibinfo{year}{2004}).

\bibitem[{\citenamefont{Huggins et~al.}(2019)\citenamefont{Huggins, McClean,
  Rubin, Jiang, Wiebe, Whaley, and Babbush}}]{huggins2019efficient}
\bibinfo{author}{\bibfnamefont{W.~J.} \bibnamefont{Huggins}},
  \bibinfo{author}{\bibfnamefont{J.}~\bibnamefont{McClean}},
  \bibinfo{author}{\bibfnamefont{N.}~\bibnamefont{Rubin}},
  \bibinfo{author}{\bibfnamefont{Z.}~\bibnamefont{Jiang}},
  \bibinfo{author}{\bibfnamefont{N.}~\bibnamefont{Wiebe}},
  \bibinfo{author}{\bibfnamefont{K.~B.} \bibnamefont{Whaley}},
  \bibnamefont{and} \bibinfo{author}{\bibfnamefont{R.}~\bibnamefont{Babbush}},
  \bibinfo{journal}{arXiv preprint arXiv:1907.13117}  (\bibinfo{year}{2019}).

\bibitem[{\citenamefont{Maslov}(2017)}]{MaslovCircuitCompIT2017}
\bibinfo{author}{\bibfnamefont{D.}~\bibnamefont{Maslov}}, \bibinfo{journal}{New
  J. Phys.} \textbf{\bibinfo{volume}{19}}, \bibinfo{pages}{023035}
  (\bibinfo{year}{2017}).

\bibitem[{\citenamefont{Pllaha et~al.}(2020)\citenamefont{Pllaha, Rengaswamy,
  Tirkkonen, and Calderbank}}]{pllaha2020unweyling}
\bibinfo{author}{\bibfnamefont{T.}~\bibnamefont{Pllaha}},
  \bibinfo{author}{\bibfnamefont{N.}~\bibnamefont{Rengaswamy}},
  \bibinfo{author}{\bibfnamefont{O.}~\bibnamefont{Tirkkonen}},
  \bibnamefont{and}
  \bibinfo{author}{\bibfnamefont{R.}~\bibnamefont{Calderbank}},
  \bibinfo{journal}{arXiv preprint arXiv:2006.14040}  (\bibinfo{year}{2020}).

\end{thebibliography}

\end{document}